\documentstyle[preprint,aps]{revtex}
\begin{document}
\newcommand\beq{\begin{equation}}
\newcommand\eeq{\end{equation}}
\newcommand\bea{\begin{eqnarray}}
\newcommand\eea{\end{eqnarray}}                         
\draft
\tightenlines
\narrowtext 
  
\title{Exact ground state number fluctuations of trapped ideal and interacting fermions}
\author{Muoi N. Tran}

\address
{Department of Physics and Astronomy, McMaster University\\
 Hamilton, Ont. L8S 4M1, Canada}
\date{\today}
\maketitle

\begin{abstract} 
We consider a small and fixed number of fermions in an isolated one-dimensional trap (microcanonical ensemble).  The ground state of the system is defined at $T=0$, with the lowest single-particle levels occupied.  The number of particles in this ground state fluctuates as a function of excitation energy.  By breaking up the energy spectrum into particle and hole sectors, and mapping the problem onto the classic number partitioning theory, we formulate a new method to calculate the {\em exact} particle number fluctuation more efficiently than the direct combinatorics method.  The  {\em exact} ground state number fluctuation for particles interacting via an inverse-square pair-wise interaction is also calculated.
\end{abstract}

\section{Introduction}
Recent experimental observation of quantum degeneracy in a dilute gas of trapped fermionic atoms at low temperatures by DeMarco and Jin \cite{demarco} inspired our previous work on fermions \cite{tranmurbhad}.  We considered an ideal gas of spinless fermions in an isolated harmonic trap in one- and two-dimensions.  In this work we restrict our study to a system of fermi gas in one-dimension.  In addition to reporting some new results we obtained for a system of ideal fermi gas, we now introduce inter-particle interaction.  Since our system is isolated (microcanonical ensemble), the results given in this paper for both ideal and interacting fermi gases are {\em exact}. 

Consider first a system containing $N$ non-interacting particles in a one-dimensional harmonic confinement, either bosons or (spinless) fermions.  The canonical partition function of this system is given by
\beq
Z_N(x)=x^{E_N(0)} \prod_{j=1}^N \frac{1}{(1-x^j)}=x^{E_N(0)}\sum_{n=0}^\infty \Omega(n,N)x^n,
\label{pf}
\eeq
where $x=e^{-\beta}$, $\beta=1/k_BT$, $E_N(0)$ is the ground state energy of the system.  $E_N(0)=N/2$ for bosons and $E_N(0)=N^2/2$ for fermions.  For simplicity we have set $\hbar \omega=1$, where $\omega$ is the oscillator frequency.  Note that the expansion coefficient $\Omega(n,N)$ is the same for bosons and fermions, and so is the canonical partition function, apart from the prefactor involving the ground state energy $E_N(0)$.  The ground state of the system is defined at $T = 0$.  For fermions the ground state consists of a set of $N$ occupied energy levels, while for bosons it is the single lowest energy level.  Given an excitation energy $n$, there are many ways in which the particles can share this energy and excite to the higher states, so that the particle number in the ground state fluctuates.  This ground state number fluctuation for ideal particles may be calculated exactly using combinatorics.  This was done by us in Ref.~\cite{tranmurbhad} for ideal fermions.  The desired quantity is the microcanonical multiplicity $\omega(n, N_{ex}, N)$, defined as the number of ways to distribute {\em exactly} $N_{ex}$ out of $N$ particles from the ground state to the excited states, given $n$ quanta of energy in a system of $N$ particles. Once $\omega(n,N_{ex},N)$ is found, the corresponding probability $P(n,N_{ex},N)$ and the {\em exact} ground state number fluctuation can be calculated straightforwardly:
\bea
P(n,N_{ex},N) & = & \frac{\omega(n,N_{ex},N)}{\sum_{N_{ex}=1}^N \omega(n,N_{ex},N)}, \label{prob} \\
\langle N_{ex} \rangle & = & \sum_{N_{ex}=1}^N P(n,N_{ex},N) N_{ex},\label{moment1} \\
\langle N_{ex}^2 \rangle & = & \sum_{N_{ex}=1}^N P(n,N_{ex},N) N_{ex}^2, \label{moment2}\\
(\delta N_0)^2 &=& \langle N_{ex}^2 \rangle - \langle N_{ex} \rangle ^2 \nonumber\\
               &=& \langle N_0^2 \rangle - \langle N_0 \rangle ^2. 
\label{fluc}
\eea
Note that by definition $P(n,N_{ex},N)$ is normalized to unity.  Eqs.~(\ref{prob})-(\ref{fluc}) apply to any system of particles, either bosons or fermions.  Note, however, that the microcanonical multiplicity $\omega(n,N_{ex},N)$ is different for different types of particles.  For fermions, apart from the fact that the ground state consists of more than one energy levels, the distribution of $N_{ex}$ particles must comply with the Pauli Principle.  For clarity, therefore, we shall attach a bracketed superscript $\omega^{\{B\}}(n,N_{ex},N)$ for bosons, and $\omega^{\{F\}}(n,N_{ex},N)$ for fermions in what follows.

Although the combinatorial method may be used to determine the microcanonical multiplicities $\omega^{B}(n,N_{ex},N)$ and $\omega^{F}(n,N_{ex},N)$ for both ideal bose and fermi gases {\em exactly}, it is very time-consuming computationally.  In the case of an ideal boson gas in a one-dimensional harmonic trap, the problem is greatly simplified due to its connection to number partitioning theory.  This is because the energy spectrum of a one-dimensional harmonic trap is equally spaced, and the $N_{ex}$ bosons are excited from {\em a single lowest energy level}. Given $n$ excitation quanta imparted to the system, the number of ways to excite {\em exactly} $N_{ex}$ bosons to the excited states, $\omega^{B}(n, N_{ex}, N)$, is equivalent to the number of ways of partitioning an integer $n$ into {\em exactly} $N_{ex}$ parts.  It is related to the expansion coefficient $\Omega(n,N)$ of the canonical partition function $Z_N(x)$ via \cite{rademacher}:
\bea
\omega^{B}(n,N_{ex},N) &=& \Omega(n-N_{ex},N_{ex}), ~ n \geq N_{ex}~\nonumber\\
                       &=&  0, \rule{1.1in}{0in}  ~otherwise.
\label{shortcutbose}
\eea
Note that Eq.~(\ref{shortcutbose}) implies that $\omega(n,N_{ex},N)=\omega(n,N_{ex},N^\prime),N^\prime \geq N_{ex}$, i.e.~independent of system size. In the language of number theory, the canonical multiplicity $\Omega(n,N)$ is the number of ways of partitioning $n$ into $1,2,3,...,N$ parts.  Clearly,
\beq
\Omega(n,N)=\sum_{N_{ex}=1}^N \omega^{B}(n,N_{ex},N).
\label{sumomegab}
\eeq
 Since the canonical multiplicity $\Omega(n,N)$ is the same for bosons and fermions (Eq.~(\ref{pf})), the following must be true:
\beq
\Omega(n,N)=\sum_{N_{ex}=1}^N \omega^{F}(n,N_{ex},N).
\label{sumomegaf}
\eeq
So even though the bosonic $\omega^{B}(n,N_{ex},N)$ is very different from the fermionic $\omega^{F}(n,N_{ex},N)$, their sum $\Omega(n,N)$ is the same.  For bosons, expression (\ref{shortcutbose}) implies that knowing the {\em canonical} multiplicity $\Omega(n,N)$ is sufficient to study the system at a microcanonical level.  It therefore allows an easy way to calculate the {\em exact} number fluctuation, instead of the more cumbersome combinatorics formulae.  This is not the case for fermions, however, since Eq.~(\ref{shortcutbose}) does not apply. This is because the (quantum mechanical) ground state of fermions consists of $N$ occupied energy levels, unlike bosons whose ground state consists of only a single lowest energy level.  To see why this multi-level ground state of fermionic system prevents the application of number theory, consider a system containing $N=2$ particles.  For simplicity take $n=2$ and consider exciting one particle, i.e., $N_{ex}=1$.  In the language of number theory, this means partitioning the integer $n=2$ into one part. There is only one way of doing this.  For bosons there is also only one way of taking $1$ out of $N=2$ particles from the ground state, and put it in the second level above, i.e., $\omega^{B}(2,1,2)=1$.  For fermions, however, either the particle which lies in the fermi level $E_F$ may get excited to the second level above, or the particle which lies just below $E_F$ gets excited to the first level above.  Thus $\omega^{F}(2,1,2)=2$, which is not equal to one as in the case of bosons.  Note that $n=2$ may also be partitioned into two parts, and $\Omega(2,2)=2$ since $2=2,1+1$.  For bosons this means exciting both particles, i.e. $\omega^{B}(2,2,2)=1$, whereas for fermions $\omega^{F}(2,2,2)=0$ since it takes at least $4$ quanta to excite $2$ fermions.  In either case the microcanonical multiplicities sum up to $2$ (Eqs.~(\ref{sumomegab})and (\ref{sumomegaf})).  So the multi-level ground state of fermionic system compensates for the restriction in the number of ways the particles may be distributed imposed by the Pauli Principle.  Fortunately, the mapping of the system of $N$ fermions in a one-dimensional harmonic trap to number partitioning theory is not impossible.  In section \ref{idealfermions}, we show that by looking at the excited energy levels and the multi-level ground state separately, we are able to connect each of these two spaces to a similar bosonic problem, which in turn pertains to number partitioning.  We successfully derive a formula similar Eq.~(\ref{shortcutbose}) which applies for fermions.  Apart from mathematical interest, the formula allows us to calculate the {\em exact} ground state number fluctuation more efficiently than the combinatorics method.

In the conventional treatment of the ideal bose gas, the number fluctuation is calculated using the grand canonical ensemble (GCE).  The result is given by \cite{kittel}
\beq
\left<(\delta N)\right> \equiv \left<(\delta N)^2\right>^{1/2}=\left[\sum_k^\infty\left<n_k\right>(1+\left<n_k\right>)\right]^{1/2}.
\label{bosefluc}
\eeq
where $\left<n_k\right>$ is the occupancy of level $k$.  As the temperature approaches zero, the particles condense to the ground state.  Relation (\ref{bosefluc}) thus yields $\left<(\delta N_0)\right> \approx \left<N_0\right>\approx \left<N\right>$.  This result is unphysical since as $T \rightarrow 0$, there is no energy to excite particles and hence the number fluctuation should vanish. This is a failure of the GCE in describing the fluctuation of an ideal bose gas at low temperatures.  Experimentally, the system has a fixed total number of particles $N$ confined in space by a trapping potential, so that it is more accurately described by a canonical ensemble (CE), or better still, a microcanonical ensemble (MCE).  This has been shown by several authors; some used property (\ref{shortcutbose}) for the MCE, and the ground state number fluctuation has been found to vanish as $T \rightarrow 0$ \cite{flucpapers1}. Within the GCE framework, however, the divergence can be removed by introducing inter-particle interactions \cite{landau}. Consider an ideal gas in a box, and recall that in GCE the particle number fluctuation is related to the thermal compressibility via
\beq
\chi_{_T}=-\frac{1}{V}\left(\frac{\partial V}{\partial P}\right)=\frac{V}{T}\left(\frac{\delta N}{N}\right)^2.
\eeq
For $T \rightarrow 0$, the pressure is independent of volume, so that $\chi_{_T} \rightarrow \infty$ and hence $\delta N \rightarrow \infty$.  With a repulsive interaction, however weak, there is a pressure due to interaction to ensure the fluctuation to be finite.  In the case of fermions, there exists a Pauli pressure even if the gas is ideal so that the ground state fluctuation finite \cite{kittel}:
\beq
\left<(\delta N_0)^2\right>^{1/2}=\left[\sum_k\left<n_k\right>(1-\left<n_k\right>)\right]^{1/2},
\label{fermifluc}
\eeq
the sum $k$ runs over the ground state levels occupied at $T=0$. As $T \rightarrow 0$, $\left<n_k\right> \rightarrow 1$ for $k$ below fermi level, so that the fluctuation goes to zero as expected.  In this paper we shall use the inverse square two-body interaction  in one-dimension (the Calogero-Sutherland model, abbreviated CSM) \cite{calogero}.  This interacting model may be mapped onto the Haldane-Wu generalized exclusion statistics (also known as fractional exclusion statistics, FES) whose quasiparticles are non-interacting, characterized by a parameter $g$ \cite{haldane,murthy,ha}.  $g=0,1$ correspond to bosons, fermions respectively.  Other values of $g$ represent various degrees of 'Pauli blocking'.  The model is solvable, and the energy spectrum is exactly known.  In \cite{bhadmurtran} we calculated the fluctuation for bosons and fermions using this interaction in both GCE and CE.  For an interacting bose gas we showed that the divergence of the number fluctuation at $T=0$ in the GCE is removed. The ground state number fluctuations of an interacting gas using CE in two- and three-dimensions have been previously calculated by a number of authors for different types of interaction \cite{bhadmurtran,flucpapers2}.  Our choice of the inverse-square two-body interaction model in one-dimension has a two-fold advantage.  First, it serves as an interacting model which allows us to demonstrate that the divergence in the GCE is removed when an interaction is introduced. Second, owing to its mapping to non-interacting FES particles, the model enables us to show that the divergence is removed even for an ideal gas provided that the Pauli blocking is not zero, no matter how weak.  To the best of our knowledge, the {\em exact} fluctuation of interacting particles in the MCE has never been calculated before.  In section \ref{interactingfermions} of this paper, we report some results of the exact fluctuation of trapped particles interacting via CSM, or equivalently, of ideal FES particles.

\section{Ideal Fermions}
\label{idealfermions}
Consider $N$ spinless fermions in a one-dimensional harmonic confinement.  The ground state is defined at $T=0$ where the $N$ particles occupy the lowest energy levels up to and including the fermi level $E_F$.  Given $n$ quanta of energy, consider breaking up $n$ into two parts:
\begin{equation}
n=n_h + n_p,
\label{n}
\end{equation}
where $n_h$ is the number of quanta it takes to bring $N_{ex}$ particles to the fermi level $E_F$ (which is equivalent to the distribution of $N_{ex}$ holes to the states below and including $E_F$), and $n_p$ is the number of quanta it takes to distribute these $N_{ex}$ particles in the excited states (above $E_F$).  This effectively divides the fermionic energy levels into two sectors: the {\em particle space} above $E_F$ and the {\em hole space} below and including $E_F$.  
Let $\omega_h^{\{F\}}(n_h,N_{ex},N)$ be the number of ways to distribute $N_{ex}$ holes in the hole space, and $\omega_p^{\{F\}}(n_p,N_{ex},N)$ be the number of ways to distribute $N_{ex}$ particles in the particle space, both according to Pauli Principle, then given $n$ quanta and $N_{ex}$ particles, 
\begin{equation}
\omega^{F}(n,N_{ex},N)=\sum_{\{n_h,n_p\}}\omega_h^{\{F\}}(n_h,N_{ex},N)~\omega_p^{\{F\}}(n_p,N_{ex},N),
\label{shortomega}
\end{equation}
where the set $\{n_h,n_p\}$ satisfies Eq.~(\ref{n}) for a given $n$.  The problem now pertains to finding $\omega_h^{\{F\}}(n_h,N_{ex},N)$ and $\omega_p(^{\{F\}}n_p,N_{ex},N)$.  At first glance this seems to be more complicated than finding a single quantity $\omega^{F}(n,N_{ex},N)$. However, recall that $\omega^{F}(n,N_{ex},N)$ is the number of ways to distribute $N_{ex}$ particles above $E_F$ with respect to {\em a set of $N$} ground state energy levels.  By breaking up the fermionic energy levels into two parts we are now distributing $N_{ex}$ particles above $E_F$ and $N_{ex}$ holes below $E_F$, both with respect to a {\em single} energy level.  As we shall see shortly, this allows us to connect to number partitioning theory as in the case of bosons. Let us now look at these two spaces separately. 

\subsection{Particle space}

First consider the particle space. This space is unbounded starting from the fermi level $E_F$.  We now consider a situation in which {\it the $N_{ex}$ particles have already been taken out of the hole space, and are residing in the fermi level $E_F$ awaiting to be distributed in the particle space}.  Note that $n_p$ quanta are distributed among $N_{ex}$ fermions, {\em with respect to only one energy level} $E_F$.  The problem is now similar to the bosonic one, except the distribution of particles must comply with the Pauli Principle.  In terms of number partitioning theory, $\omega_p^{F}(n_p,N_{ex},N)$ is the number of ways of partitioning an integer $n_p$ into $N_{ex}$ {\em different} parts, with $N_{ex} \leq N$.  As in the case of bosons, it is related to the canonical multiplicity via \cite{rademacher}:
\bea
\omega_p^{\{F\}}(n_p,N_{ex},N) &=& \Omega(n_p-\bigtriangleup_p,N_{ex}), \nonumber \\
 			&=& \Omega(n_p-\frac{N_{ex}(N_{ex}+1)}{2},N_{ex}).
\label{omegap}
\eea
Remarkably, this is the same form as Eq.~(\ref{shortcutbose}) for bosons with the shifted energy $N_{ex}$ replaced by $\bigtriangleup_p$.   This shifted energy, $\bigtriangleup_p=N_{ex}(N_{ex}+1)/2$, is in fact the minimum energy it takes to excite $N_{ex}$ particles from $E_F$,
\beq
n_p^{min}(N_{ex})=\frac{(N_{ex}+1)N_{ex}}{2}.
\label{npmin}
\eeq
Note that the partition function of this space is no longer given by that of fermions.  The canonical multiplicity in Eq.~(\ref{omegap}) should therefore be thought of as the bosonic multiplicity.  In other words, the problem is now mapped onto a similar bosonic problem, with the restriction that the parts of an integer being partitioned are {\em different}.  This notion is most helpful when we discuss the hole space.

\subsection{Hole space}
We now consider taking $N_{ex}$ particles out of the multi-level ground state and put them in the fermi level $E_F$ (or equivalently, creating $N_{ex}$ holes in the ground state).  For a given $n_h$ quanta, we wish to find $\omega_h^{\{F\}}(n_h,N_{ex},N)$, the number of ways of doing this.  Unlike the particle space, the dimension of the hole space is bounded, set by the number of particle $N$ of the system.  Note that for a given number $N_{ex}$ of particles, the Hilbert space of available states for $N_{ex}$ holes is dependent on the value of $N_{ex}$ itself and is given by:
\begin{equation}
N_H=N-N_{ex}.
\end{equation} 
We need to find the partition function of this space for each $N_{ex}$, and derive a formula similar to Eq.~(\ref{omegap}) for $\omega_h^{\{F\}}(n_h,N_{ex},N)$.  This may be done by considering a new system {\em containing $N_{ex}$ bosons}, whose energy space is bounded and is given by $N_H+1$ including the ground state.  The goal is to determine $Z^{H}_{N_{ex}}(x)$, and expand this in terms of the coefficient $\Omega_h(n_h,N_{ex})$.  The $N_{ex}$-hole partition function of this hypothetical system can be found using \cite{borrmann}:
\beq 
Z^{H}_{N_{ex}}(x) = \frac{1}{N_{ex}} \sum_{j=1}^{N_{ex}}Z^{H}_1(jx) Z^{H}_{N_{ex}-j}(\beta),
\label{recursion} 
\eeq
where $Z^{H}_1(x)$ is the one-particle partition function of the system containing $N_{ex}$ bosons and is given by:  
\begin{equation}
Z^{H}_1(x)=\sum_{i=0}^{N_H}x^{i}.
\end{equation}
Note that the one-particle partition function needs to be determined for a given $N_{ex}$.  Using this, $Z^{H}_{N_{ex}}(x)$ may then be found from (\ref{recursion}).  Once  $Z^{H}_{N_{ex}}(x)$ is found, we may expand it in powers of $x$:
\beq
Z^{H}_{N_{ex}}(x) = \sum_i\Omega_h(i,N_{ex}) x^i.
\eeq
We are now ready to determine a formula for $\omega_h^{\{F\}}(n_h,N_{ex},N)$.  Because the hole space includes $E_F$, the minimum energy to create a hole (or dig a particle) to put in the fermi level is zero, since there already is a particle there; for two holes the minimum energy is one, for three holes it is three...etc. In general, 
\beq
n_h^{min}(N_{ex}) = \frac{(N_{ex}-1)N_{ex}}{2}.
\label{nhmin}
\eeq
Similar to Eq.~(\ref{omegap}) with the energy shift $\bigtriangleup_h$ given by (\ref{nhmin}), the number of ways of creating $N_{ex}$ holes in the hole space $\omega^{F}_h(n_h,N_{ex},N)$ is given by: 
\beq
\omega_h(n_h,N_{ex},N)=\Omega_h(n_h-\frac{N_{ex}(N_{ex}-1)}{2},N_{ex}).
\label{omegah}
\eeq
Using Eqs.~(\ref{omegap}) and (\ref{omegah}), the number of ways of distributing $N_{ex}$ fermions to the excited states, Eq.~(\ref{shortomega}), now reads:
\beq
\omega^{F}(n,N_{ex},N) = \sum_{\{n_h,n_p\}}\Omega(n_p-\frac{N_{ex}(N_{ex}+1)}{2},N_{ex})~\Omega_h(n_h-\frac{N_{ex}(N_{ex}-1)}{2},N_{ex}).
\label{shortcutfermi}
\eeq

It is obvious that if there is no hole space, $\omega_h(n_h,N_{ex},N)=\Omega_h(n_h-\bigtriangleup_h,N_{ex})=1$, the sum over the set $\{n_h,n_p\}$ vanishes since $n=n_p$, the energy shift $\bigtriangleup_p=n_p^{min}(N_{ex})=N_{ex}$, and Eq.~(\ref{shortomega}) reduces to Eq.~(\ref{shortcutbose}) for bosons. 

Using Eqs.~(\ref{shortcutfermi}) and (\ref{prob})-(\ref{fluc}), we calculated the ground state number fluctuation of fermions for $N=100$.  The result is shown in Fig.~\ref{figure1}. For comparison we also show the corresponding result in the CE.

\section{Ideal FES particles}
\label{interactingfermions}
We now calculate the {\em exact} ground state fluctuation of particles interacting via CSM, or equivalently, non-interacting particles obeying FES \cite{calogero,haldane,murthy,ha}.  The Hamiltonian of the CSM is given by \cite{calogero}:
\begin{equation}
H = \sum_{i=1}^N \left [ -\frac{\hbar^2}{2m} \frac{\partial^2}{\partial 
x_i^2} + \frac{1}{2} m\omega^2 x_i^2 \right ] +
\frac{\hbar^2}{m} \sum_{i<j=1}^N \frac{g(g-1)}{(x_i-x_j)^2}
\label{ham}
\end{equation}
with the dimensionless coupling parameter $g \geq 0$. The particles are 
confined in a harmonic well and the thermodynamic limit
is obtained by taking $\omega \rightarrow 0$ as $N\rightarrow \infty$, with 
$\omega N=constant$. In the thermodynamic limit, the properties of the system 
are translationally invariant, and would be the same if the particles were 
on a line, or a circle, instead of a harmonic confinement. To make the 
problem well-defined quantum mechanically, we have to demand that the wave functions go to zero as $|x_i-x_j|^g$ whenever two particles i and j approach 
each other. Since the particles cannot cross each other, we may choose the 
wave function to be either symmetric (bosonic) or antisymmetric (fermionic). 
For $g=0$ and $1$, the model describes free bosons and free fermions respectively.  

In \cite{bhadmurtran} we have calculated the fluctuation for FES particles in the GCE and CE.  To do this in the MCE, we need to find $\omega^{\{g\}}(n,N_{ex},N)$ for a given $g$.  The energy spectrum of the CSM Hamiltonian (\ref{ham}) is exactly known, and may be expressed in terms of the quasiparticle energy spectrum $\epsilon_g$ of non-interacting FES particles, which is given by \cite{murthy}:
\beq
\epsilon_g=\epsilon_k - (1-g)N_k,
\label{quasien}
\eeq
with $\hbar\omega \equiv 1$.  $N_k$ is defined as the number of particles below energy level $k$, $\epsilon_k=(k-\frac{1}{2}),k=1,2,...$ is the harmonic oscillator energy spectrum.
The canonical N-particle partition function of FES particles reads:
\bea
Z_N &=& x^{E^{\{g\}}_N(0)}\prod_{j=1}^N \frac{1}{(1-x^j)}, ~\nonumber \\
    &=& x^{E^{\{g\}}_N(0)}\sum_{n=0}^{\infty} \Omega(n,N) x^n.
\label{chacha}
\end{eqnarray}
Note that the canonical multiplicity $\Omega(n,N)$ is still the same for FES particles as for bosons and fermions.  The only effect of the interaction strength $g$ is to alter the overall ground state energy $E^{\{g\}}_N(0)=gN(N-1)/2 + N/2$ which reduces to that of fermions, bosons for $g=1, 0$ respectively.  

Using Eq.~(\ref{quasien}), one may determine the spectra of FES particles for any value of $g$.  In the Fig.~(\ref{figure2}) we show an example of how to find the spectrum of semions ($g=1/2$) for some values of excitation quanta $n$ and number of particles $N=5$. Starting from the fermionic spectrum on the left, for each value of $n$, the spectrum of semions is drawn and the resulting particles which are in the excited states are determined.  In the case of $n=3$, for instance, the only possibility for fermions is to excite one particle and there are 3 ways of doing this as shown.  Therefore, $\omega^{\{F\}}(3,1,5)=3$, and $\omega^{\{F\}}(3,2,5)=\omega^{\{F\}}(3,3,5)=0$ since $3$ quanta is too few to excite $2$ or more particles.  For semions, however, $\omega^{\{1/2\}}(3,1,5)=1$, $\omega^{\{1/2\}}(3,2,5)=2$, and $\omega^{\{1/2\}}(3,3,5)=0$.  Note that $\Omega(3,5)=3$ in both cases.  Following the same procedure, we found the microcanonical multiplicities $\omega^{\{g\}}(n,N_{ex},N)$ for $g=3/4,1/2,1/4$, $n=1...16$ quanta, and $N=2...5$ particles.  The method gets more cumbersome, however, for larger values of $n$ and $N$.  Ideally, one wishes to be able to determine $\omega^{\{g\}}(n,N_{ex},N)$ from $\omega^{\{F\}}(n,N_{ex},N)$ or $\omega^{\{B\}}(n,N_{ex},N)$, or from $\Omega(n,N_{ex})$ similar to Eqs.~(\ref{shortcutbose}) and (\ref{shortcutfermi}).  This general formula for $\omega^{\{g\}}(n,N_{ex},N)$ for any $g$, if it exists, is yet to be found.  Here, we report the finding of the microcanonical multiplicities for only two values of $g$: $g=\frac{N-2}{N-1}$ (close to fermions), and $g=\frac{1}{N-1}$ (close to bosons).  For $g=\frac{N-2}{N-1}$, $\omega^{\{\frac{N-2}{N-1}\}}(n,N_{ex},N)$ is found to be given by: 
\beq
\omega^{\{\frac{N-2}{N-1}\}}(n,N_{ex},N) = \omega^{\{F\}}(n+N,N_{ex},N)-\omega^{\{F\}}(n+N,N_{ex},N-1),
\label{omegagf}
\eeq
For $g=\frac{1}{N-1}$:
\bea
\omega^{\{\frac{1}{N-1}\}}(n,N_{ex},N)\rule{0.18in}{0in} &=& \omega^{\{B\}}(n,N_{ex},N), \rule{1in}{0in} N_{ex} \neq N-1,N \nonumber \\
\omega^{\{\frac{1}{N-1}\}}(n,N-1,N) &=&\omega^{\{B\}}(n,N-1,N)+\omega^{\{B\}}(n-1,N-1,N), \nonumber \\
\omega^{\{\frac{1}{N-1}\}}(n,N,N)\rule{0.3in}{0in}  &=&\omega^{\{B\}}(n-N,N,N).
\label{omegagb}
\eea
These multiplicities must, of course, satisfy
\beq
\Omega(n,N)=\sum_{N_{ex}=1}^N \omega^{\{g\}}(n,N_{ex},N).
\label{property}
\eeq
For $N \leq 5$ and $n \leq 16$, the values found using Eq.~(\ref{omegagf})-(\ref{omegagb}) were verified with those found using the direct combinatorial method as described above.

Although a general formula for the microcanonical multiplicities for any $g$ has not been found, an important point concerning them is observed.  For a given number of particles $N$, consider a set of discrete values of $g$, 
\beq
g = 1, \frac{N-2}{N-1}, \frac{N-3}{N-1},..., \frac{1}{N-1}, 0. 
\label{g}
\eeq
Since the levels of the FES particles are shifted by an amount which depends on the value of $g$ and the number of particles below (see Eq.~(\ref{quasien})), for some values of $g$ a particle might lie very close to the last level of the ground state, $E^{\{g\}}_F$. However, the particles are considered excited if and only if they lie {\em above} $E^{\{g\}}_F$, no matter how close.  This results in the multiplicities $\omega^{\{g\}}(n,N_{ex},N)$, for $\frac{N-i}{N-1} > g \geq \frac{N-(i+1)}{N-1}, i=1,...,N-1 $ {\em to be the same}.  Note the equal sign in $g \geq \frac{N-(i+1)}{N-1}$. For instance, 
\bea
\omega^{\{g\}}(n,N_{ex},N) &=& \omega^{\{B\}}(n,N_{ex},N),~ for~ \frac{1}{N-1} > g \geq 0. \nonumber
\eea
Note that for these values of $g$ given by Eq.~(\ref{g}), the 'Fermi level' $E^{\{g\}}_F=N-(1-g)(N-1)$ is integral.  So for $g=(N-k)/(N-1)$, where $k$ is an integer, the 'Fermi level' lines up with the $(k-1)^{th}$ level below $E_F$ of fermions.  In box 1, $n=0$ of Fig.~\ref{figure2}, for instance, where $N=5$ and $g=2/4$ ($k=3$), the 'Fermi level' lines up with the second level below $E_F$.  This explains why the microcanonical multiplicities for some range of $g$ such that the 'Fermi level' lies between the $k^{th}$ and $(k+1)^{th}$ levels of fermions are the same.  This is due the discrete nature of the energy levels. 




\section{discussion}
In our previous work (Ref.~\cite{tranmurbhad}) where the direct combinatorial method was used, the fermionic calculation of the ground state fluctuation was restricted to a low number of particles $N$ and quanta $n$ using a normal office computer (Pentium III, 500 cpu).  For a relatively small number of particles (e.g., $N=10$), at higher excitation $n$ the combinatorics method is more time-consuming due to the rapid increase in the number of possibilities with $n$.  The method described in section \ref{idealfermions} translates the problem in combinatorics into the problem in calculating the partition functions of the hole space, the latter being simpler computationally.  Although this method is still time-consuming and the calculation for larger $N$ ($N \geq 100$) is still not possible using our office computer, it is more effective for higher number of quanta and relatively small number of particles.  For demonstration we display the ground state fluctuation of fermions as a function of energy quanta $n$ in Fig.~\ref{figure1} for $N=30$.  We also show the corresponding curve in CE, which is the same as GCE except at very low temperature (see  e.g., Ref.~\cite{tranmurbhad}) for comparison.  As expected, both go to zero as $T \rightarrow 0$, with the MCE fluctuation less than that given by CE for all $n$.  Note that the two fluctuations are very different, even for very high excitations.  At $n=6000$ which is $200 \times E_F$, the CE curve still differs from the MCE one by about $14\%$.  It was shown in Ref.~\cite{tranmurbhad} for $N=15$ that the ground state occupancy $\left<N_0\right>=\sum_k^{E_F}\left<n_k\right>$ for the two ensembles are very similar.  Clearly, the number fluctuation is more sensitive to the ensemble used.  Therefore, for a relatively small particle number, while it may be adequate to use the CE (or GCE) to describe a thermodynamic quantity such as the ground state occupancy, it should be used with caution when calculating the number fluctuation and related quantities.

Using Eqs.~(\ref{omegagf}), (\ref{omegagb}) and (\ref{prob})-(\ref{fluc}), we calculated the {\em exact} ground state number fluctuation for interacting particles.  In Fig.~\ref{figure3}a we display the fluctuation for $N=5$, $g=3/4,1/4$.  For comparison we also show the fluctuation of free fermions and bosons.  The values for $g=3/4,1/4$ obtained using the direct combinatorial method (see Fig.~\ref{figure2}) for $n \leq 16$ are also shown.  Note that the curves agree with these values exactly.  Note also that the number fluctuation for free fermions and free bosons cross at a certain energy, with the fermionic one starting from smaller at small quanta, to larger at high quanta.  This is because the number of possibilities of creating holes within the fermi sea and distributing particles above, which starts from low at small energy, increases more rapidly than for bosons whose ground state consists of only one level.  Similar behaviours are observed for FES particles, whose $g$ values ($g=3/4,1/4$) represent partial Pauli blocking which are both less than that of fermions. 

Eqs.~(\ref{omegagf}) and (\ref{omegagb}) in principle may be applied for any $N$.  However, since they involve the addition of two quantities which may be very large at large quanta and number of particles, there is a difficulty in obtaining their values accurately.  Therefore, without lost of accuracy, we restricted the calculations to $N=10$.  Fig.~\ref{figure3}b shows the ground state number fluctuation for $N=10$, $g=1,8/9,1/9,0$.  Note that the curves for $g=8/9,1/9$ are closer to those of fermions and bosons than for $g=3/4,1/4$, $N=5$, in Fig.~\ref{figure3}a.  As $N$ gets larger we expect the fluctuation curves for $g=\frac{N-2}{N-1}$ and $\frac{1}{N-1}$ to come very close to those of free femions and free bosons.  The formulae (\ref{omegagf}) and (\ref{omegagb}) are therefore useful only for systems with a small number of particles.  Note also that in both graphs, the results for $g=\frac{N-2}{N-1}$ are closer to those of fermions than for $g=\frac{1}{N-1}$ to bosons.  This is can be understood from comparing Eqs.~(\ref{omegagf}) and (\ref{omegagb}).  Eq.~(\ref{omegagf}) involves the {\em difference} of the fermionic microcanonical multiplicities of two system sizes, whose values may be very similar especially at low quanta.  Eq.~(\ref{omegagb}), however, involves the {\em addition} of two bosonic microcanonical multiplicities.  This brings the microcanonical multiplicities of $g= \frac{1}{N-1}$ further from those of bosons than $g= \frac{N-2}{N-1}$ from those of fermions. 
    
Finally, In Figs.~\ref{figure4}a,b we compare these exact fluctuation curves with the corresponding CE curves for $g= \frac{N-2}{N-1},\frac{1}{N-1}$, $N=5,10$.    From \cite{bhadmurtran}, the expression for the canonical ground state fluctuation is given by:
\beq
\left<(\delta N_0)^2\right>^{\{g\}} = g((\delta N_0)^2)^F + (1-g)((\delta N_0)^2)^B
\label{cefluctg}
\eeq
Clearly, the results between the two ensembles are very different especially for small $n$.  In both graphs the microcanonical fluctuations are less than the canonical ones for all $n$.  Note that the CE formula (\ref{cefluctg}) allows one to calculate the fluctuation for a general $g$.  The determination of a general formula for the microcanonical multiplicity, $\omega^{\{g\}}(n,N_{ex},N)$, and hence the ground state number fluctuation for any $g$, if it exists, remains a challenge.

\section{acknowledgments}
MNT would like to acknowledge the Natural Sciences and Engineering Research Council (NSERC) of Canada for financial assistance.  Special thanks to  Dr. J.~C.~Waddington for suggesting an important point regarding ideal fermions,  Dr. D.~Sen for going through the exact combinatorial results of FES particles, and  Dr's. M.~V.~N.~Murthy, R.~K.~Bhaduri for guidance and a careful reading of the manuscript.

\newpage
\begin{figure} 
\caption{Ground state number fluctuation of fermions as a function of excitation energy $n$ (in unit of $\hbar\omega$) for $N=30$.  The result in the CE is also shown for comparison.}
\label{figure1}
\vspace{7 mm}

\caption{Spectra of semions ($g=1/2$) derived from those of fermions using Eq.~(\ref{quasien}).  The (quantum mechanical) ground state is defined at $n=0$.  The spectra for different energies are separated by boxes. Within a box there may be more than one configuration, each of which contains the spectrum of ideal fermions on the left and semions on the right.  The number of excited particles, $N_{ex}$, are indicated in each case.}
\label{figure2}
\vspace{7 mm}

\caption{(a) Ground state number fluctuation curves as functions of excitation quanta $n$ for $N=5$, $g=1, 3/4, 1/4, 0$.  For clarity the low energy part is shown in the inset.  The data represented by the symbols are obtained using the direct combinatorial method as shown in Fig.~\ref{figure2}; $\bigcirc$ for $g=3/4$, and $\Box$ for $g=1/4$.  (b) Same as in (a), for $N=10$ and $g=1, 8/9, 1/9, 0$.} 
\label{figure3}
\vspace{7 mm}

\caption{(a) Comparison between the {\em exact} fluctuations as discussed in this paper with those in the CE (Eq.~(\ref{cefluctg})) for $N=5$, $g=3/4,1/4$.
(b) Same as in (a), $N=10$, $g=8/9,1/9$.} 
\vspace{7 mm}
\label{figure4}
\end{figure}

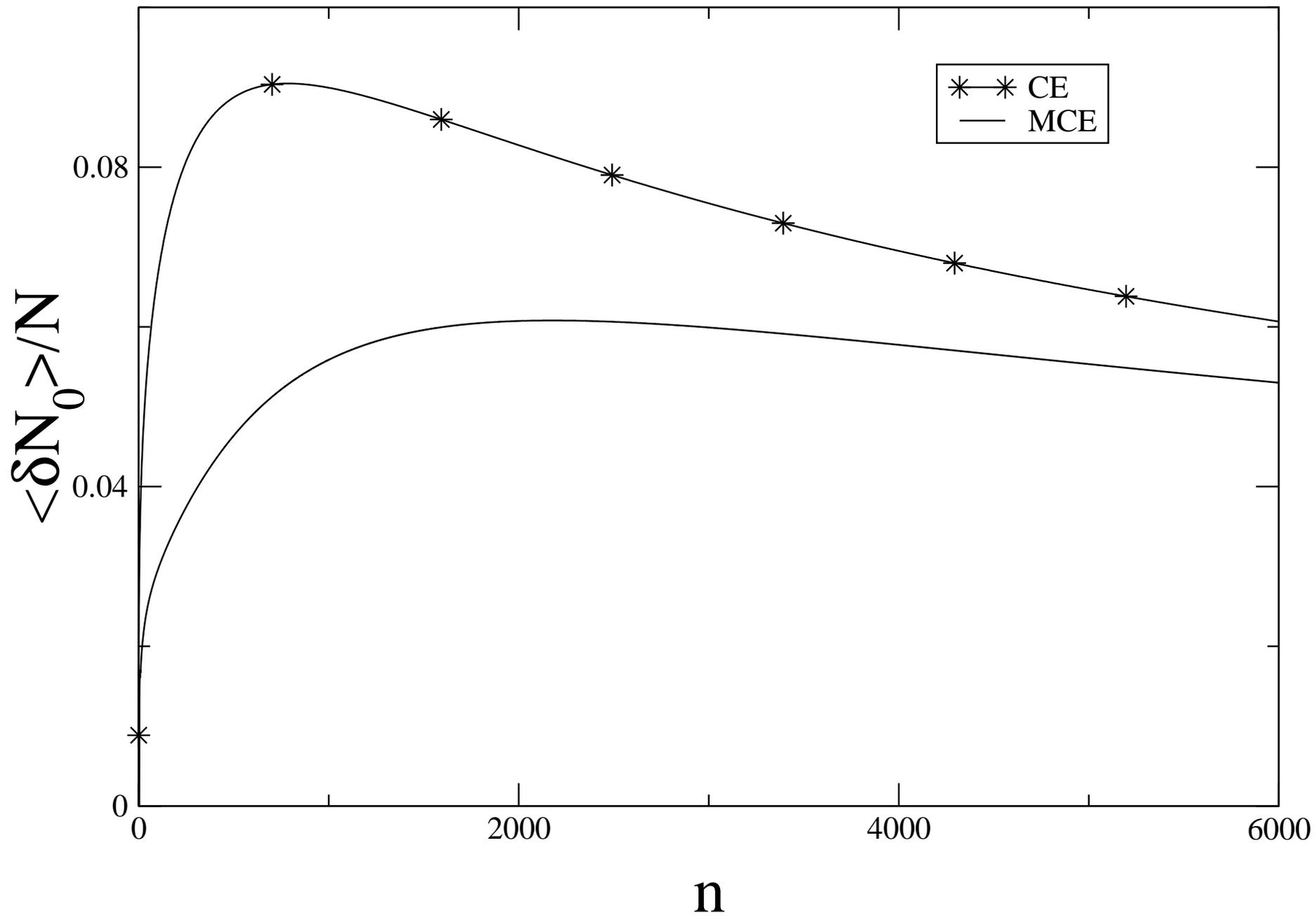
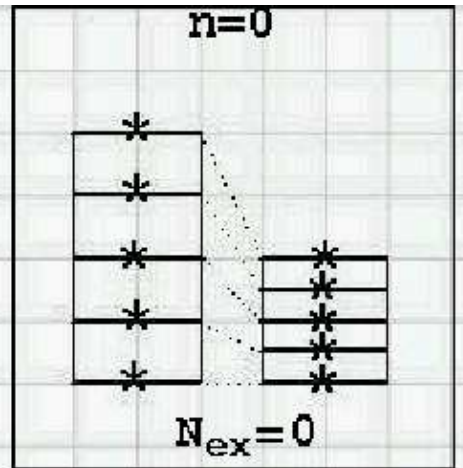
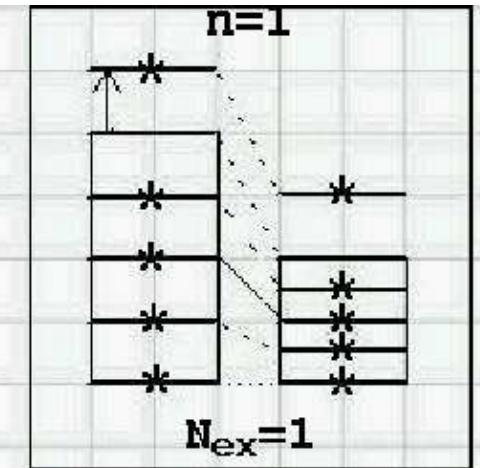
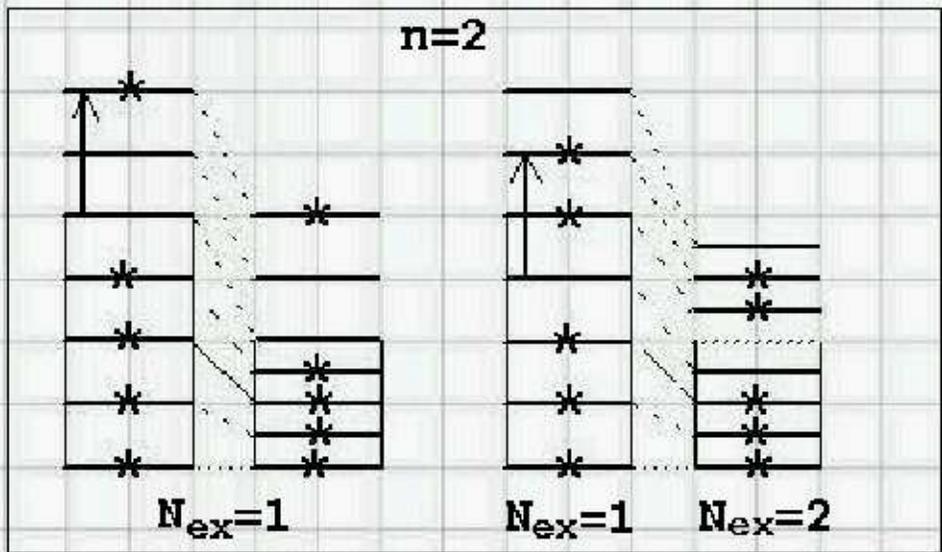
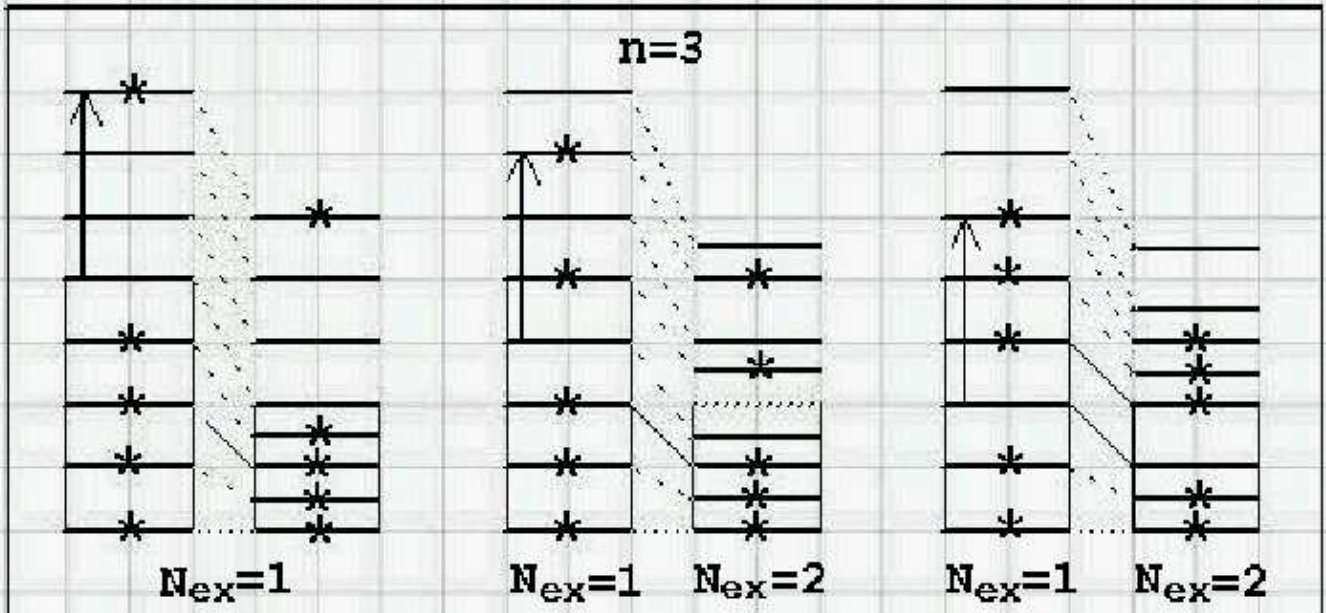
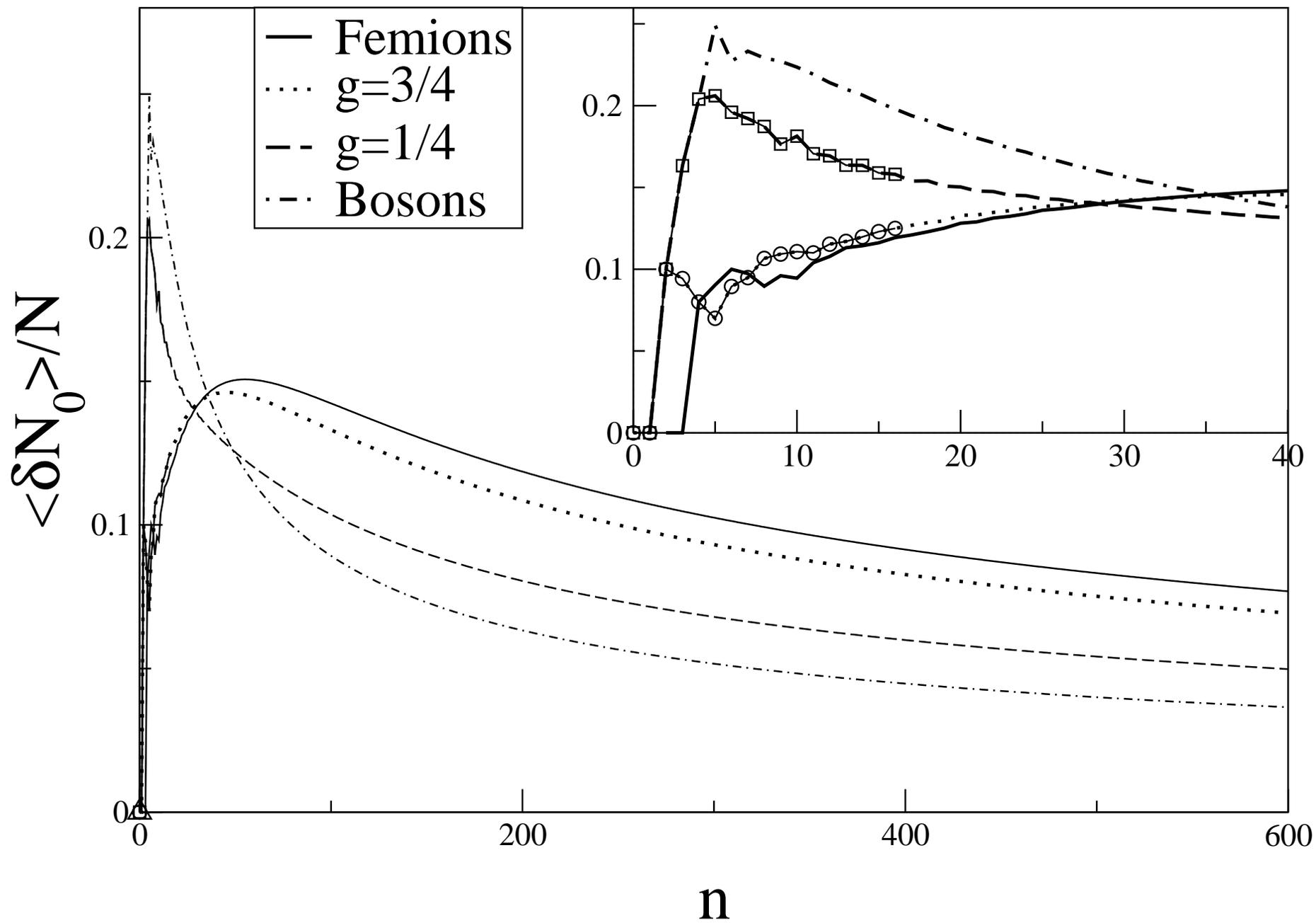
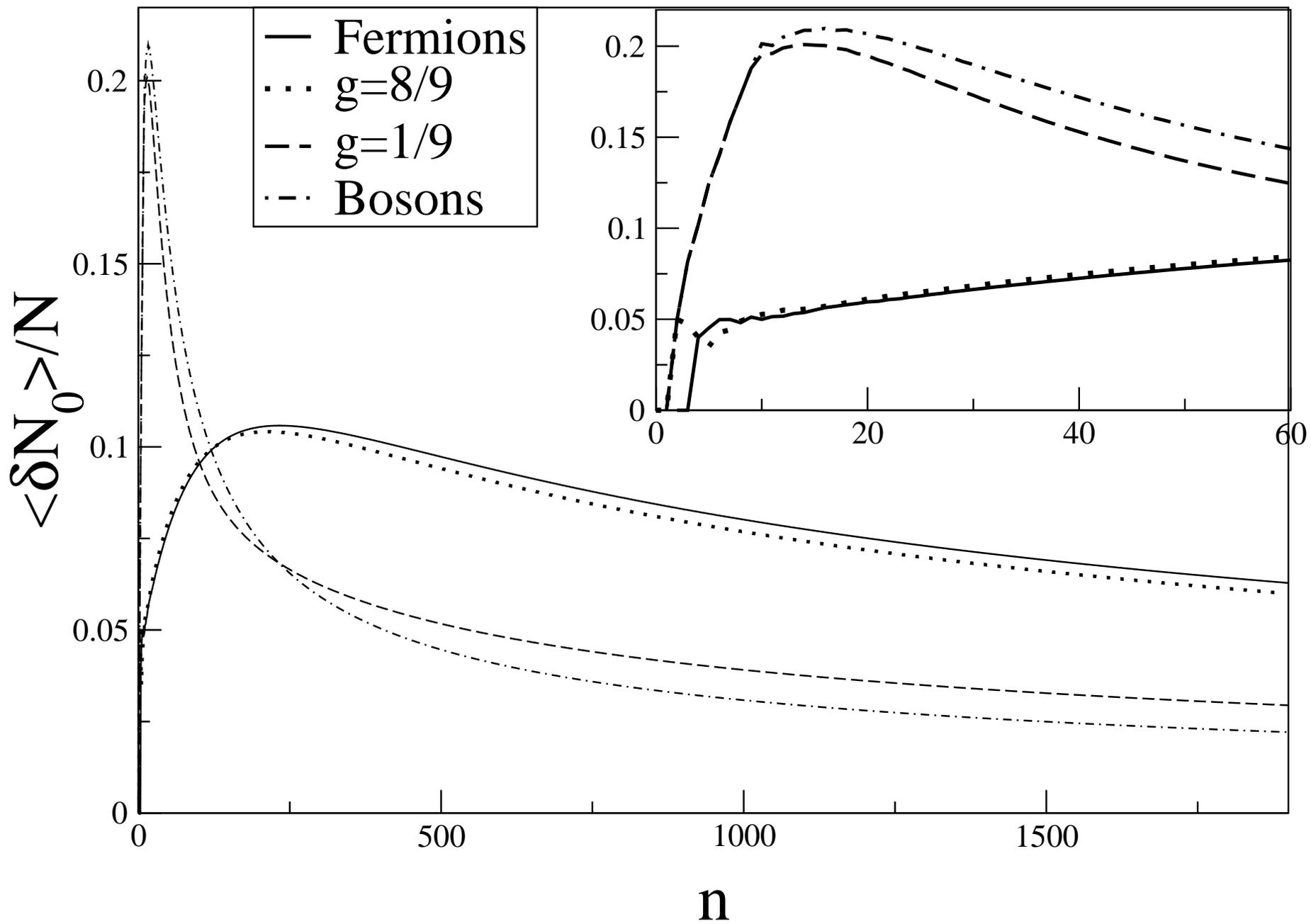
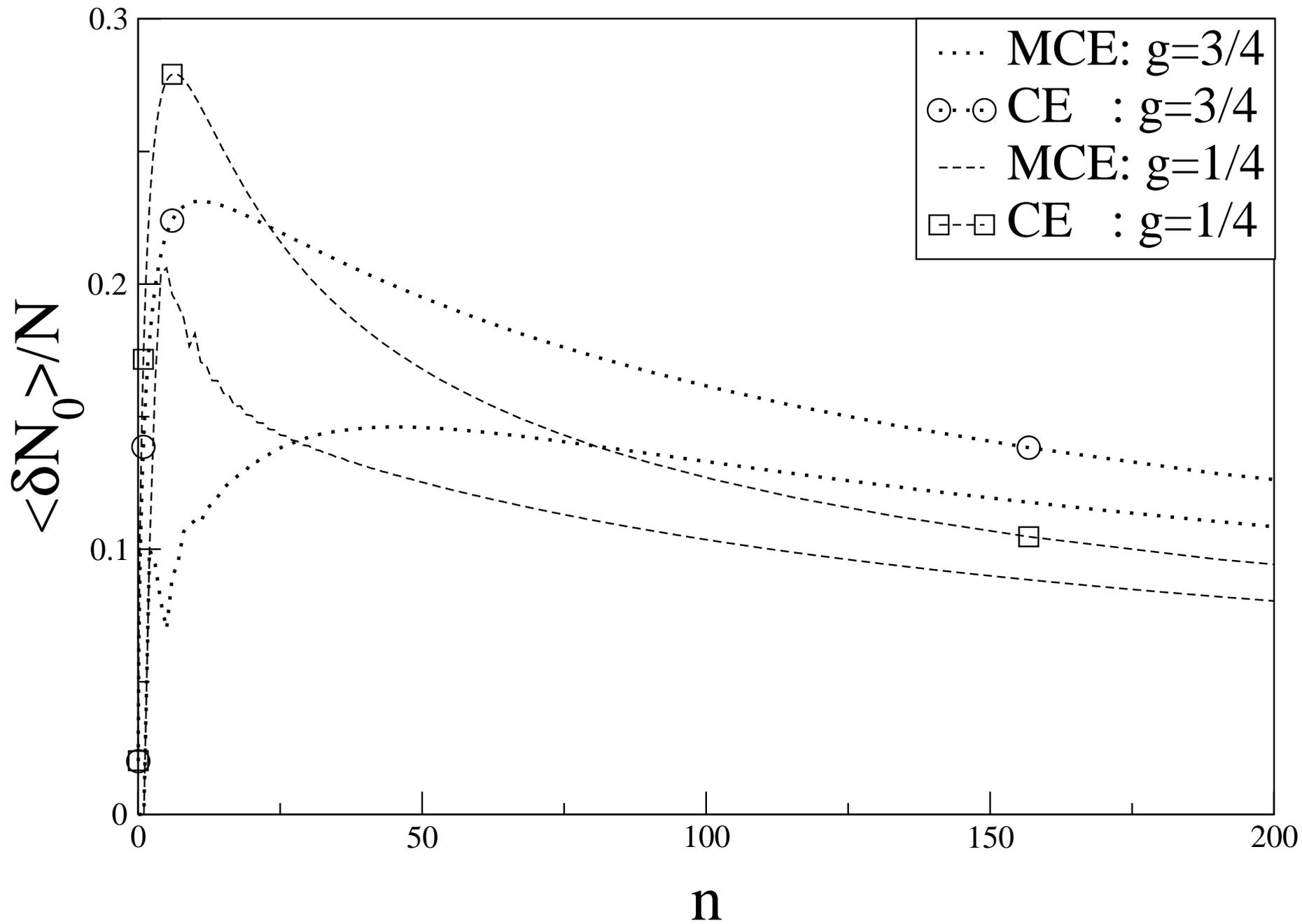
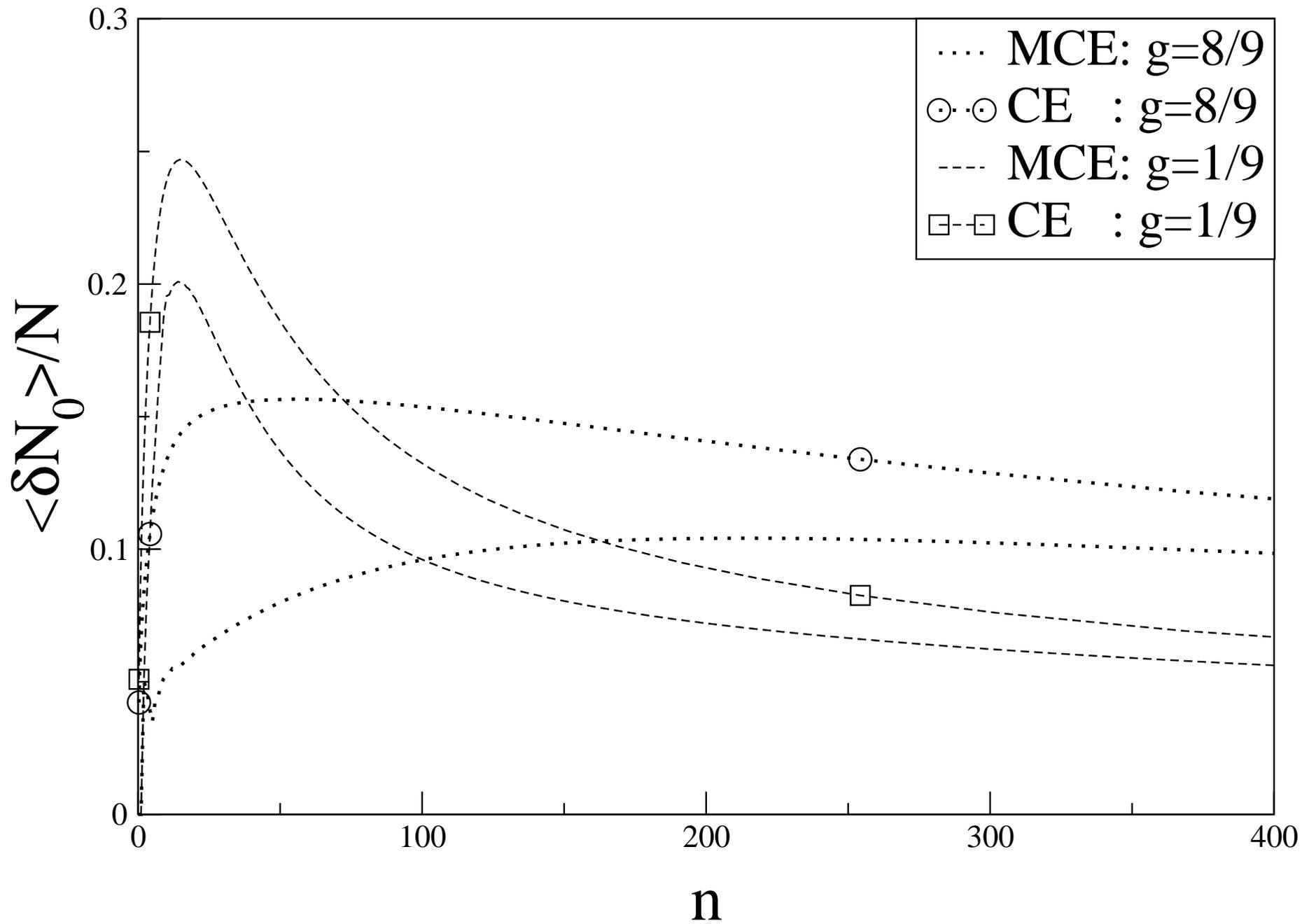
\end{document}